\documentclass[prb,twocolumn,letter, showpacs]{revtex4}
\usepackage{amssymb}
\usepackage{epsfig}
\usepackage{epstopdf}
\usepackage{amsmath}
\usepackage{bbm}
\usepackage{graphicx}

\begin{document}

\author{Luca de' Medici$^{1,3}$}
\author{Xin Wang$^2$}
\author{Massimo Capone$^{3,4}$}
\author{Andrew J. Millis$^2$}
\affiliation{ $^{(1)}$Department of Physics and Center for Materials Theory, Rutgers the State University of NJ,
136 Frelinghuysen Road, Piscataway, NJ 08854 and Laboratoire de Physique des Solides, Universit\'e Paris-Sud, CNRS, UMR 8502, F-91405 Orsay Cedex, France \\
$^{(2)}$Department of Physics, Columbia University, 538 W. 120$^{th}$ Street, New York, NY 10027 USA\\
$^{(3)}$Dipartimento di Fisica,
Universit\`a di Roma ``La Sapienza'', Piazzale A. Moro 2, I-00185,
Rome, Italy\\
$^{(4)}$SMC Center, CNR-INFM, Piazzale A. Moro 2, I-00185 and ISC-CNR, Via dei Taurini 18, I-00185, Rome, Italy \\}
\title{Correlation Strength, Gaps and Particle-Hole Asymmetry in  High-$T_c$ Cuprates: a Dynamical Mean Field Study of the Three-Band Copper-Oxide Model}

\date{\today}

\hyphenation{}

\begin{abstract}
The three-band model relevant to high temperature copper-oxide
superconductors is solved using single-site dynamical mean field
theory and a tight-binding parametrization of the copper and oxygen
bands.  For a band filling of one hole per unit cell the
metal/charge-transfer-insulator phase diagram is determined.  The
electron spectral function, optical conductivity and quasiparticle
mass enhancement are computed as functions of electron and hole
doping for parameters such that the corresponding to the
paramagnetic metal and charge-transfer insulator sides of the one
hole per cell phase diagram.  The optical conductivity is computed
using the Peierls phase approximation for the optical matrix
elements. The calculation includes the physics of ``Zhang-Rice
singlets''.   The effects of antiferromagnetism on the magnitude of
the gap and the relation between correlation strength and
doping-induced changes in state density are  determined.  Three band
and one band models are compared. The two models are found to yield
quantitatively consistent results for all energies less than about
$4$eV,  including energies in the vicinity of the charge-transfer
gap. Parameters on the insulating side of the metal/charge-transfer
insulator phase boundary lead to gaps which are too large and
near-gap conductivities which are too small relative to data. The
results place the cuprates clearly in the intermediate correlation
regime, on the paramagnetic metal side of the metal/charge-transfer
insulator phase boundary.
\end{abstract}


\maketitle

\section{Introduction}

The interplay of band theoretic issues of hybridization and chemical
bonding with  the quantum chemical issue of strong local
correlations is basic to the physics of many important materials. In
an important paper \cite{Zaanen85} Zaanen, Sawatzky and Allen
classified insulating transition metal oxides  as
``charge-transfer'' or ``Mott'' insulators according to whether the
physics could be discussed solely in terms of strong correlated
transition metal  $d$ states or whether transitions to O $2p$ states
were important to the low energy physics.  The issue arises with
particular force in the high temperature superconductors where the
Cu $d^9$ and $d^{10}$ states are not far in energy from O $2p$
states, but the Cu $d^8$ state is very far away in energy. In this
circumstance a strong particle-hole asymmetry is expected
\cite{Emery87}, with  doped electrons residing on Cu sites whereas
doped holes reside mainly on the O, but may be bound to Cu spins
creating ``Zhang-Rice singlets".\cite{Zhang88}

Quantifying this appealing physical picture requires solving an
electronic structure problem with multiple scales, including  a
correlation energy on the Cu site $\sim8-10$eV, a Cu-O energy level
difference of 2-4eV,\cite{Mila88,Veenendal94} and a Cu-O
hybridization $\sim1.6$eV.\cite{Andersen95} In this paper we use
single-site dynamical mean-field theory (DMFT)  to solve a model
involving both copper and oxygen orbitals,  developing a
comprehensive  theoretical picture of the electronic structure and
optical conductivity of undoped and doped cuprate materials across
the charge-transfer-insulator to charge-transfer metal phase
diagram, including the effect of antiferromagnetism on the spectra
and optics. We use newly improved Exact Diagonalization
(ED)\cite{Caffarel94,Capone04} and Continuous-Time  Quantum Monte
Carlo (CT-QMC) \cite{Werner06} impurity solvers. These methods have
different sources of error and we find consistent results with the
two methods.

Our work is related to previous work of Dopf {\sl et
al.},\cite{Dopf92} Georges {\sl et al.},\cite{Georges93} and Z\"olfl
{\sl et al.} \cite{Zolfl98}  who each studied one particular
parameter value. Modern developments in computers and solvers mean
that we are able to obtain much more information. Our work also has
some overlap with more recent work of Macridin {\sl et al.}
\cite{Macridin05} who used the Dynamical Cluster Approximation on a
four site cluster to study momentum dependence and the onset of
superconductivity. We compare our findings to very recent work of
Weber {\sl et al.}\cite{Weber08} who used similar methods to study a
similar model. Weber {\sl et al.} focused on specific parameters; we
focus on the spectral functions and conductivities  over a wider
energy range, varying the charge-transfer gap to explore all regions
of the theoretical phase diagram and present a comparison of the low
energy behavior of the copper-oxygen model to that of  an effective
one-band model.

We present evidence that a one-band model provides a reliable
picture of the spectral functions and conductivity for frequencies
less than about $4$eV (note that this range extends about a factor
of two above the charge-transfer gap in frequency).  We find an
electron-hole asymmetry in the self-energy. The asymmetry is much
more pronounced for parameters such that the undoped material is a
charge-transfer insulator (insulating even in the paramagnetic
phase). The asymmetry, however, is not reflected in the Fermi
velocity renormalization or the low frequency optical matrix
oscillator strength, where changes in the electronic structure
compensate for the differences in correlation strength.  Comparison
of our results to data suggests that the cuprates are on the
metallic side of the single-site DMFT phase diagram, with
antiferromagnetism being essential to produce the gap in the undoped
material.

The rest of this paper is organized as following: in section
\ref{mode} we describe the model, in section \ref{resu} we present
the calculated phase diagram, spectral functions, self-energies and
optical conductivities, in section \ref{comparison} we compare the
copper-oxygen model results to those obtained from computations
performed on the one-band Hubbard model. Section \ref{conc} contains
a summary of our results and a conclusion.

\section{Model}\label{mode}

We analyze the canonical two dimensional ``copper-oxygen"
Hamiltonian \cite{Emery87,Varma87} retaining the Cu $d_{x^2-y^2}$
and O $2p_\sigma$ orbitals (whose momentum ($p$) components are
created by the operators $d^{\dagger}_{p\sigma},
p^{\dagger}_{x,p\sigma}, p^{\dagger}_{y,p\sigma}$).  We allow for
the possibility of two-sublattice antiferromagnetism by doubling the
unit cell. The Hamiltonian is therefore a 6-band model $H=H_{\rm
6band}+H_{\rm int}$ in the magnetic Brillouin zone. To write the
band theoretical part we divide the lattice into two sublattices,
$A$ and $B$, distinguish the oxygen sites displaced from the Cu in
the $x$ and $y$ directions, adopt the basis
$|\psi\rangle=\left(d^{\dagger}_{Ap\sigma} ,
p^{\dagger}_{A,x,p\sigma},
p^{\dagger}_{A,y,p\sigma},d^{\dagger}_{Bp\sigma},
p^{\dagger}_{B,x,p\sigma}, p^{\dagger}_{B,y,p\sigma}\right)$ and
write
\begin{equation}
H_{\rm 6band}=\left(\begin{array}{cc}H_A & H_M \\H_M & H_B\end{array}\right)
\label{H6band}
\end{equation}
where the $3\times3$ matrices are
\begin{eqnarray}
H_A&=&H_B=\left(\begin{array}{ccc}
\varepsilon_d & t_{pd}e^{i\frac{p_x}{2} }& t_{pd}e^{i\frac{p_y}{2} }\\
t_{pd}e^{-i\frac{p_x}{2} } & \varepsilon_p & 0 \\
t_{pd}e^{-i\frac{p_y}{2} } & 0 & \varepsilon_p
\end{array}\right),\\
H_M&=&\left(\begin{array}{ccc}
0 &- t_{pd}e^{-i\frac{p_x}{2} }&- t_{pd}e^{-i\frac{p_y}{2} }\\
-t_{pd}e^{i\frac{p_x}{2} } & 0 & 0 \\
-t_{pd}e^{i\frac{p_y}{2} } & 0 & 0
\end{array}\right)
\end{eqnarray}
and $H_{\rm int}=U\sum_i n_{d\uparrow}n_{d\downarrow}$. Here we
neglect oxygen-oxygen hopping. We use the value $t_{pd}=1.6$eV
suggested by band theory calculations.\cite{Andersen95}

Because only the Cu site is interacting, we may integrate out the
oxygen band to obtain an effective one-orbital model,  which we
solve in the single-site DMFT \cite{Georges96} using
ED\cite{Caffarel94,Capone04} and CT-QMC\cite{Werner06} methods
described in the literature. In the ED calculations typically $9$
bath states were used and the results were verified by occasional
large calculations; for CT-QMC temperatures studied were typically
$T=t_{pd}/25$ for the phase diagram and $T=t_{pd}/16$ for spectral
functions and conductivities.  The main result of the DMFT
calculation is a self-energy which, in the antiferromagnetic phase,
is spin dependent and takes the form
\begin{equation}
{\mathbf \Sigma}_\sigma(z)=\left(\begin{array}{cccccc}\Sigma_{A,\sigma}(z) & 0 & 0 & 0 & 0 & 0 \\0 & 0 & 0 & 0 & 0 & 0 \\0 & 0 & 0 & 0 & 0 & 0 \\0& 0 & 0 & \Sigma_{B,\sigma}(z) & 0 & 0 \\0 & 0 & 0 & 0 & 0 & 0 \\0 & 0 & 0 & 0 & 0 & 0\end{array}\right)
\end{equation}
with $\Sigma_{A,\uparrow}=\Sigma_{B,\downarrow}$.

In the single-site DMFT method the self-energy is momentum
independent so the conductivity may be computed from \cite{Millis05}
\begin{eqnarray}
\sigma(\Omega)&=&2\int_{-\infty}^\infty \frac{d\omega}{\pi}\int \frac{d^2p}{(2\pi)^2}\frac{f(\omega)-f(\omega+\Omega)}{\Omega}
\\
&\times&{\mathrm{Tr}}\left[{\bf j}_{\rm 6band}(p){\bf A}(\omega+\Omega,p){\bf j}_{\rm 6band}(p){\bf A}(\omega,p)\right],
\nonumber
\label{sigmamatrix}
\end{eqnarray}
where the integral is over the magnetic Brillouin zone, the  spectral function ${\mathbf A}$ is
\begin{equation}
{\bf A}(\omega,p)=\frac{i}{2}\left({\bf
G}(\omega,\Sigma(\omega+i\epsilon,p)-{\bf
G}(\omega,\Sigma(\omega-i\epsilon,p)\right),
\end{equation}
the matrix Green function ${\bf G}$ at frequency  $z$ and chemical
potential $\mu$ is $ {\bf G}(z, p)=\left(z{\bf 1}+\mu-{\bf
\Sigma}(z)-H_{\rm 6band}\right)^{-1}$, the current operator ${\bf
j}=\delta {\bf H}/\delta p_x$ is
\begin{equation}
{\bf j}_{\rm 6band}= i\frac{t_{pd}}{2}\left(\begin{array}{cccccc}0 &
e^{i\frac{p_x}{2}} & 0 & 0 & e^{-i\frac{p_x}{2}} & 0
\\-e^{-i\frac{p_x}{2}} & 0 & 0 & -e^{i\frac{p_x}{2}}  & 0 & 0 \\0 &
0 & 0 & 0 & 0 & 0 \\0 & e^{-i\frac{p_x}{2}}  & 0 & 0 &
e^{i\frac{p_x}{2}}  & 0 \\-e^{i\frac{p_x}{2}}  & 0 & 0 &
-e^{-i\frac{p_x}{2}}  & 0 & 0 \\0 & 0 & 0 & 0 & 0 &
0\end{array}\right), \label{j6band}
\end{equation}
and $f(\omega)$ is the Fermi function.

For comparison we have studied the one-band model described by
$H_{\rm 1band}=H_{\rm hop}+U_{\rm
eff}\sum_in_{i\uparrow}n_{i,\downarrow}$ with
\begin{equation}
H_{\rm hop}=-2t\sum_p\left(\begin{array}{cc}0 & \cos p_x+\cos p_y
\\ \cos p_x+\cos p_y& 0\end{array}\right) \label{H1band}
\end{equation}
$t=0.37$eV is chosen to reproduce the non-interacting bandwidth of
the Cu-O antibonding band passing through the Fermi level and
$U_{\rm eff}$ fixed so as to reproduce the correlation gap.

\section{Results}\label{resu}
\subsection{Phase diagram}

\begin{figure}[t]
\includegraphics[width=0.8\columnwidth,angle=-90]{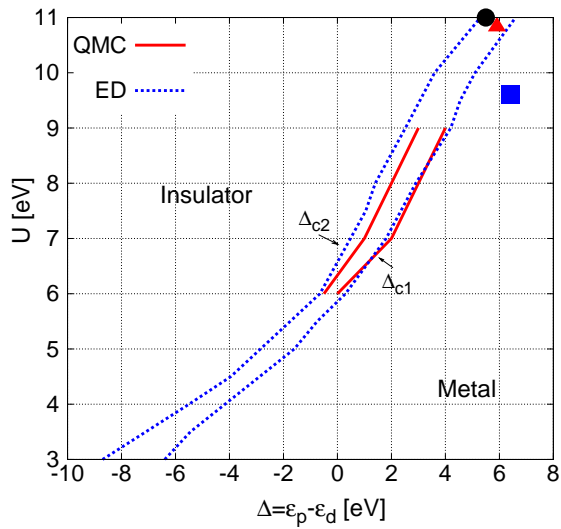}
\caption{Metal-insulator phase diagram in plane of interaction $U$
and $p-d$ level splitting $\Delta$ for one hole per CuO$_2$ unit in
paramagnetic phase. Dotted lines (blue online): phase boundaries
from ED calculation at $T=0$ ; solid lines (red online) indicate
limit of stability of metallic phase from CT-QMC calculation at
$T=1/40$eV. In the region between the two lines metallic and
insulating solutions coexist.  Square (blue online), circle (black
online) and triangle (red online):  parameters studied by
Ref.~\onlinecite{Dopf92}, Ref.~\onlinecite{Zolfl98},
Ref.~\onlinecite{Macridin05} respectively.} \label{phasediagram}
\end{figure}

Fig. \ref{phasediagram} shows the boundary between metallic and
paramagnetic insulating solutions, calculated for one hole per
CuO$_2$ unit as a function of interaction strength $U$ and Cu-O
energy level splitting $\Delta$($=\varepsilon_p-\varepsilon_d$) in
the paramagnetic phase. As in the single-site DMFT of the  one-band
Hubbard model \cite{Georges96} a coexistence region is observed
where both metallic and insulating solutions exist.  We define the
insulating and metallic boundaries of the coexistence region at
fixed large $U$ to be $\Delta_{c2}$ and $\Delta_{c1}$ respectively.
The phase diagram is obtained at $T=0$ using the ED solver; the
results were verified using CT-QMC by scanning  $\Delta$ at selected
$U$ values.  We find almost perfect agreement for $\Delta_{c1}$; the
CT-QMC calculation of $\Delta_{c2}$ line represents the smallest
$\Delta$ value at which a metallic phase can be found at temperature
$T=1/40$eV. The known T-dependence of the single-site DMFT boundary
for the single-band Hubbard model \cite{Georges96} suggests that the
discrepancy may simply be a finite-temperature effect related  to
the different temperatures used in the two calculations.

\begin{figure}[t]
\includegraphics[angle=-90,width=0.85\columnwidth]{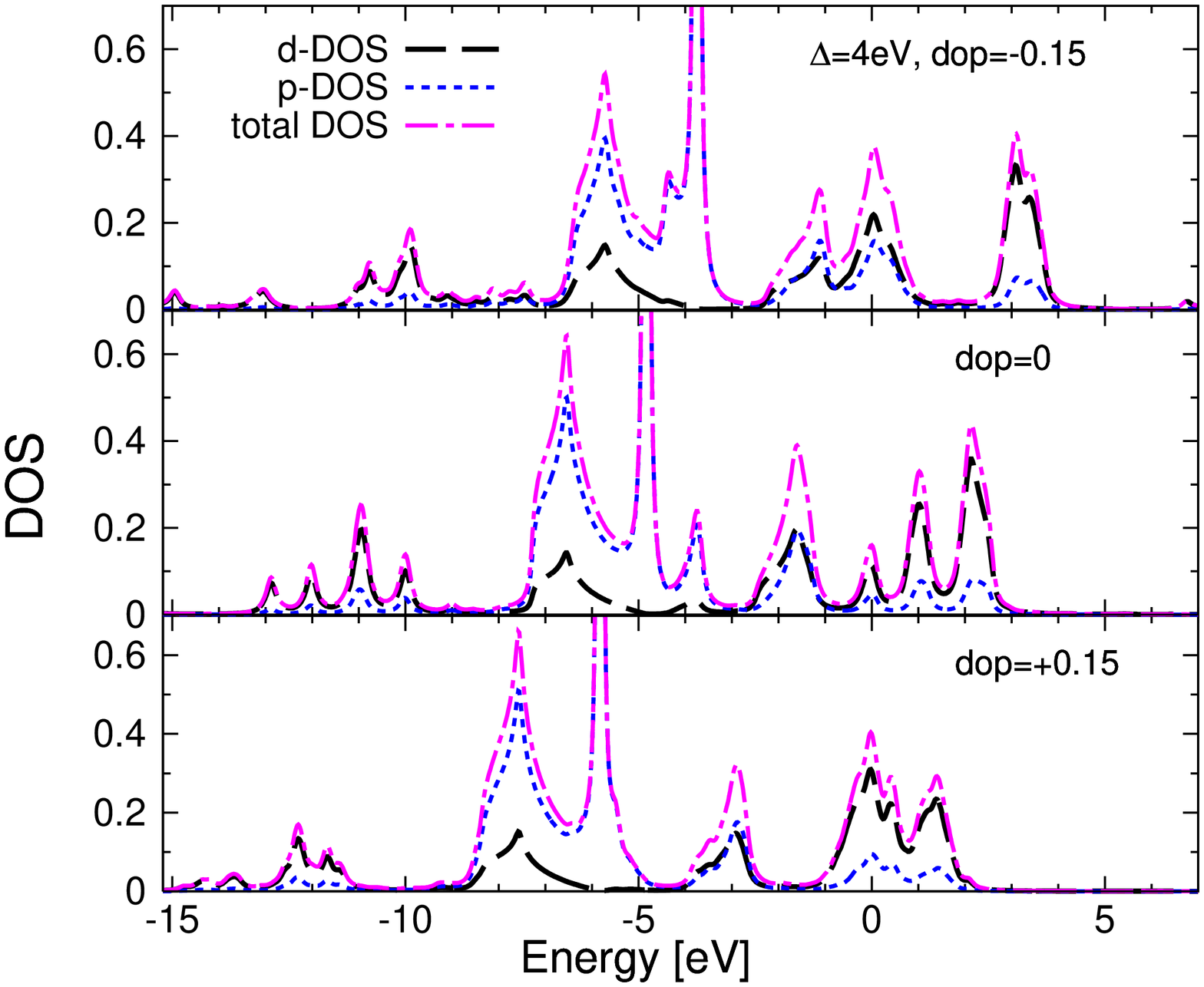}

\vspace{.3in}
\includegraphics[angle=-90,width=0.85\columnwidth]{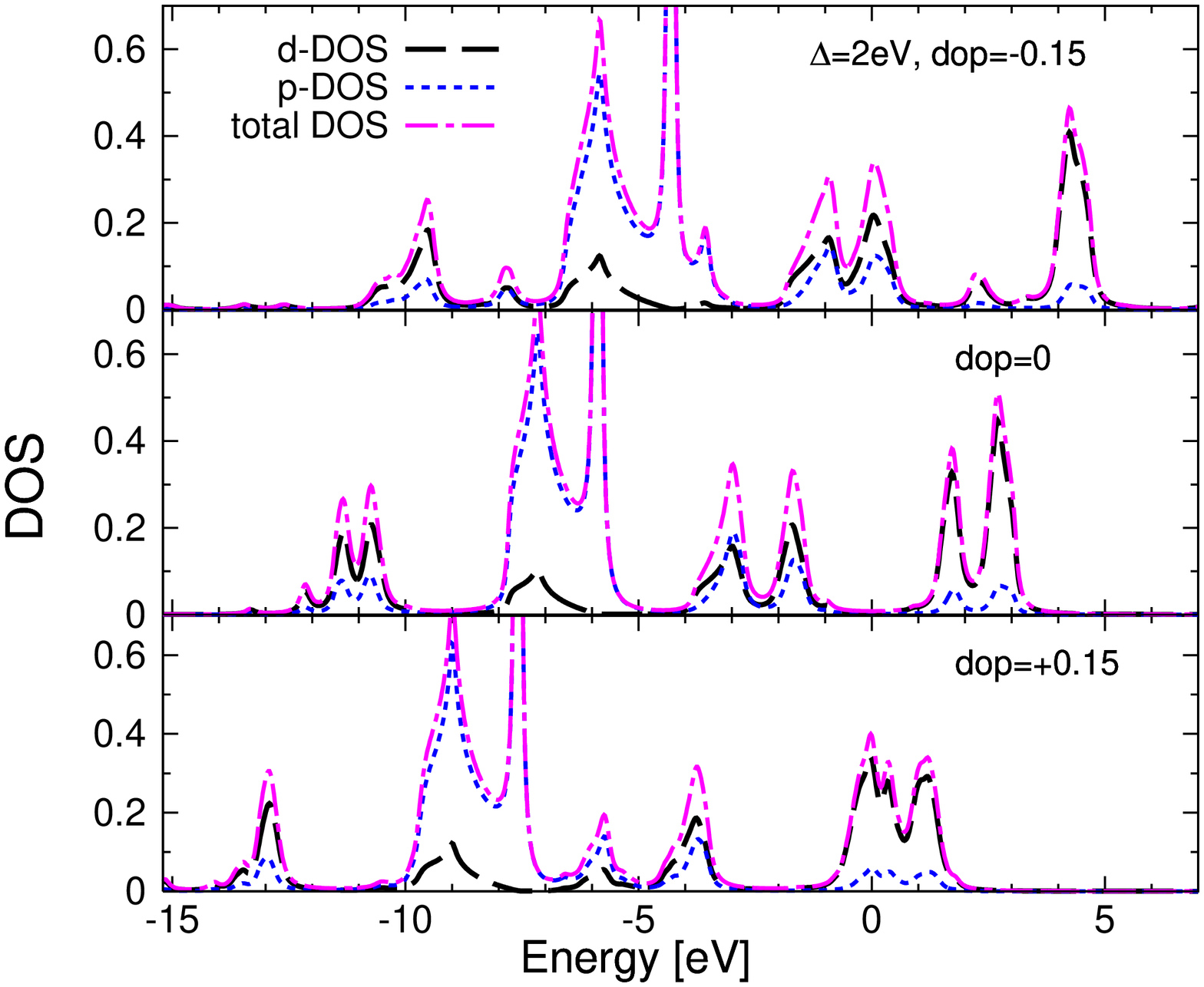}
\caption{Electron spectral function per spin for paramagnetic
three-band model (negative frequency: removal spectrum; positive
frequency: addition spectrum) (solid lines) and projections onto $d$
(dashed; red online) and $p$ (dotted line; blue online) states
calculated with the ED solver. Parameters: $U=9$eV and $\Delta=4$eV
(upper panels) and  $\Delta=2$eV (lower panels). Upper graphs:
$0.15$ hole doping ($\Delta=4$eV: $\varepsilon_d=-7.7$eV,
$\varepsilon_p=-3.7$eV; $\Delta=2$eV, $\varepsilon_d=-6.3$eV,
$\varepsilon_p=-4.3$eV) middle graphs: undoped ($\Delta=4$eV,
$\varepsilon_d=-8.8$eV $\varepsilon_p=-4.8$eV and $\Delta=2$eV,
$\varepsilon_d=-7.9$eV $\varepsilon_p=-5.9$eV). Lower graphs: $0.15$
electron doping ($\Delta=4$eV, $\varepsilon_d=-9.8$eV,
$\varepsilon_p=-5.8$eV; $\Delta=2$eV; $\varepsilon_d=-9.6$eV,
$\varepsilon_p=-7.6$eV).} \label{dos}
\end{figure}

\subsection{Spectral function}

Fig. \ref{dos} presents the many body density of states (DOS)
(electron removal spectrum for energy $<0$ and  electron addition
spectrum for energy $>0$) as well as the projections onto the Cu-$d$
and O-$2p_\sigma$  states calculated in the paramagnetic phase using
the ED method. The zero of energy is the chemical potential. The
upper panel presents results for  $\Delta>\Delta_{c2}$ (so within
single-site DMFT the paramagnetic phase of the undoped material is
metallic) while the lower panel presents results for $\Delta
\sim\Delta_{c2}$ such that the undoped material is a charge-transfer
insulator.  The main difference between the two spectra is the
presence or absence of a gap in the undoped material.

At lowest energy (binding energy $\sim11$eV) a peak is seen, of mainly $d$-character. This peak corresponds to removing one electron and leaving the Cu in the $d^8$ state; it is pushed down from the bare $d^8$ energy by a level repulsion due to hybridization with the O states. Thus it is not correct to  identify the position of this peak directly with $U$, as is sometimes done in literature.

In the binding energy range $4{\rm eV}\lesssim\varepsilon\lesssim 8{\rm eV}$ a mainly oxygen-like band is seen. The very sharp peak corresponds to the non-bonding oxygen state; it would be broadened if oxygen-oxygen hopping were taken into account (although the vanishing of the Cu-O hybridization at the $\Gamma$ point means that a singularity would remain at the bare oxygen energy).  The broad structure of mixed oxygen and copper character lying below the non-bonding state may be thought of as the  ``bonding" linear combination of Cu and O states pushed down below the non-bonding O level by hybridization to the Cu $d$ level.  This feature was identified in Ref.~\onlinecite{Weber08} as the copper upper Hubbard band corresponding to the $d^8$ state, but Ref.~\onlinecite{Weber08} did not present results over a wide enough range to determine if the  $\sim11$eV peak (which we find to  correspond to  the $d^8$ state) was present in their calculations.

The structure of mixed Cu-O character at binding energies  in the range $\varepsilon \sim 1-4 {\rm eV}$ corresponds to the Zhang-Rice singlet states. Calculations   (not shown here) in which the Cu orbital is forced to be fully spin polarized show that these states correspond to holes with the same spin as the deep-lying removal state. The states in the electron addition spectrum  are mainly of  Cu $d^{10}$ character, and play the role of the upper Hubbard band.

The effect of antiferromagnetism on the magnitude of the gap in the
charge-transfer insulator state has been the subject of debate, with
Ref.~\onlinecite{Comanac08} arguing on the basis of Hubbard model
calculations that antiferromagnetism increases the gap significantly
while the conductivity calculations presented in
Ref.~\onlinecite{Weber08} were interpreted as indicating  no
significant effect of antiferromagnetism on the gap. We have used
the ``quasiparticle equation'' method of Ref.~\onlinecite{Wang09a}
to determine the gap values at $U=9$eV, $\Delta=2$eV finding that
the gap in the paramagnetic insulating phase is $2.87$eV while the
addition of antiferromagnetism shifts the gap to $3.47$eV.  Some of
the difference between our results and those of
Ref.~\onlinecite{Weber08} may arise from the extremely small value
of the calculated paramagnetic-state conductivity in the near gap
region, especially for $\Delta<\Delta_{c2}$, which may have led
those authors to overestimate the gap in the paramagnetic state.

Upon doping, two changes occur. First, the chemical potential moves
into the Zhang-Rice band (hole doping) or the upper Hubbard band
(electron doping).  For  $\Delta\sim\Delta_{c2}$ the associated
changes in chemical potential are substantial:  the Fermi level,
measured relative to the non-bonding oxygen peak, shifts by almost
2eV; for the $\Delta\sim \Delta_{c1}$ the Fermi level changes rather
less. Second, as can be seen by inspection of the Figure~\ref{dos}
and from Table \ref{areas}, the relative strengths of the different
spectral features evolve. The issue of the number of states created
by doping has received some attention in the literature as a
signature of ``Mottness".\cite{Meinders93, Phillips04} We find that
the changes in the spectrum  do not have a universal  doping or
interaction-strength  dependence;  however in the ``charge-transfer
insulating"  regime $\Delta \lesssim \Delta_{c2}$, each doped hole
adds roughly two states  to the Zhang-Rice band and one to the upper
Hubbard band, while doping with electrons does essentially the
opposite.  In the paramagnetic metal case the changes in electronic
structure are larger.

\begin{table}
 \centering
 \begin{tabular}{lcllcl}
  &\hspace{.1in}$\Delta=4$eV&\hspace{.4in}$\Delta=2$eV\\
 \end{tabular}
 \begin{tabular}{|c||c|c|c||lc|c|c|}
 \hline dop & -0.15 & 0.00 & 0.15&& -0.15& 0.00& 0.15\\ \hline \hline $d^8$ & 0.39 & 0.53 & 0.40 && 0.50 & 0.59 & 0.41  \\
  \hline Z-R & 1.2 & 0.72 & 0.54 && 1.0 & 0.90 & 0.74 \\
 \hline UHB & 0.70 & 1.0 & 1.3 && 0.80 & 0.98 & 1.23 \\ \hline \end{tabular}
 \caption{Integrated density of states (both spins) for the $d^8$ (lowest-lying),
 ``Zhang-Rice" (ZR) and ``Upper Hubbard Band" (UHB) spectral features discussed in the text
 for $U=9$eV, $\Delta=4$eV (paramagnetic metal in undoped case) and $\Delta=2$eV (paramagnetic insulator in undoped case)
 at dopings indicated.}\label{areas}
\end{table}

\subsection{Self-energy and velocity renormalization}

\begin{figure}[t]
\includegraphics[angle=-90,width=0.85\columnwidth]{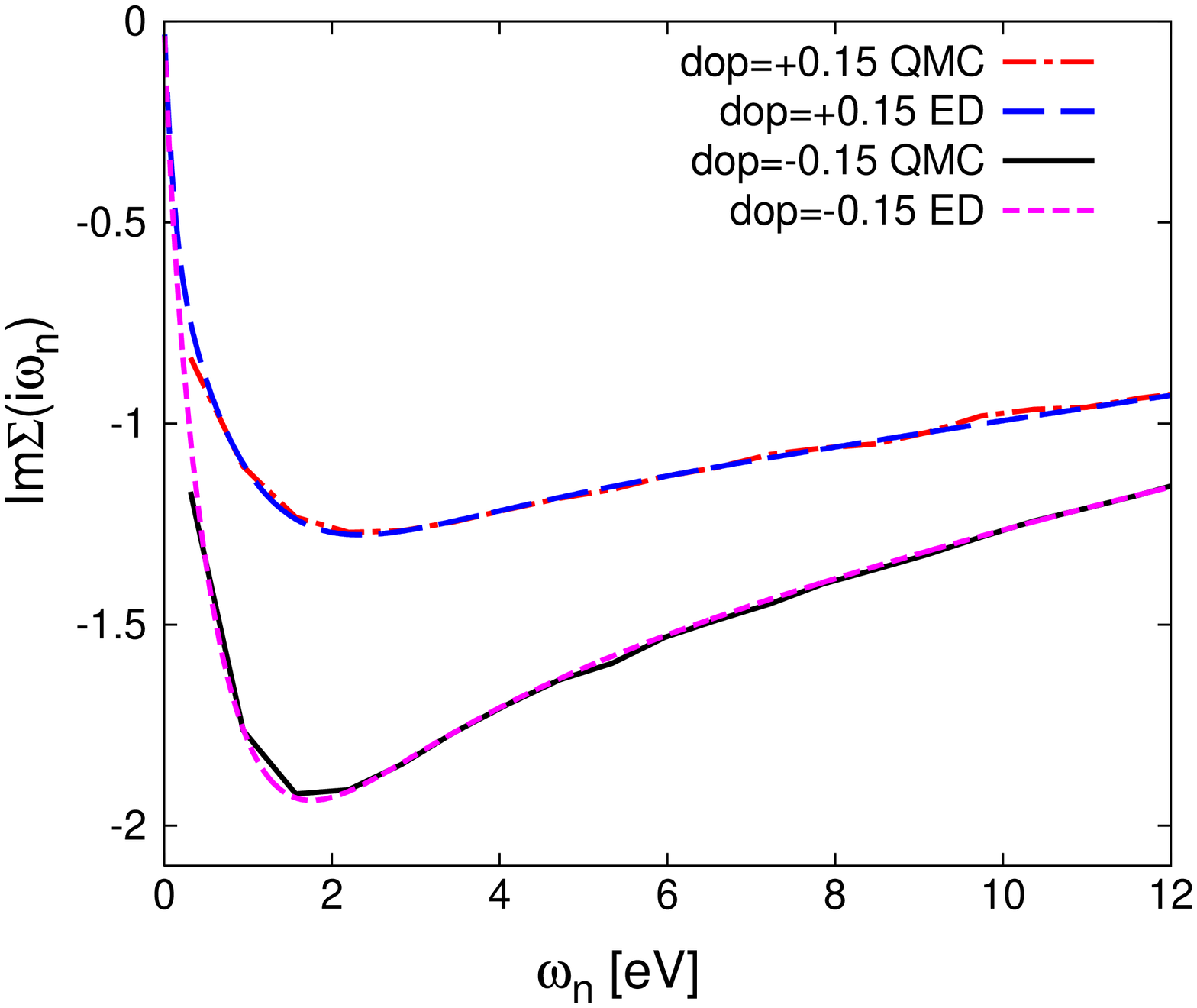}
\vspace{0.5in}
\includegraphics[angle=-90,width=0.85\columnwidth]{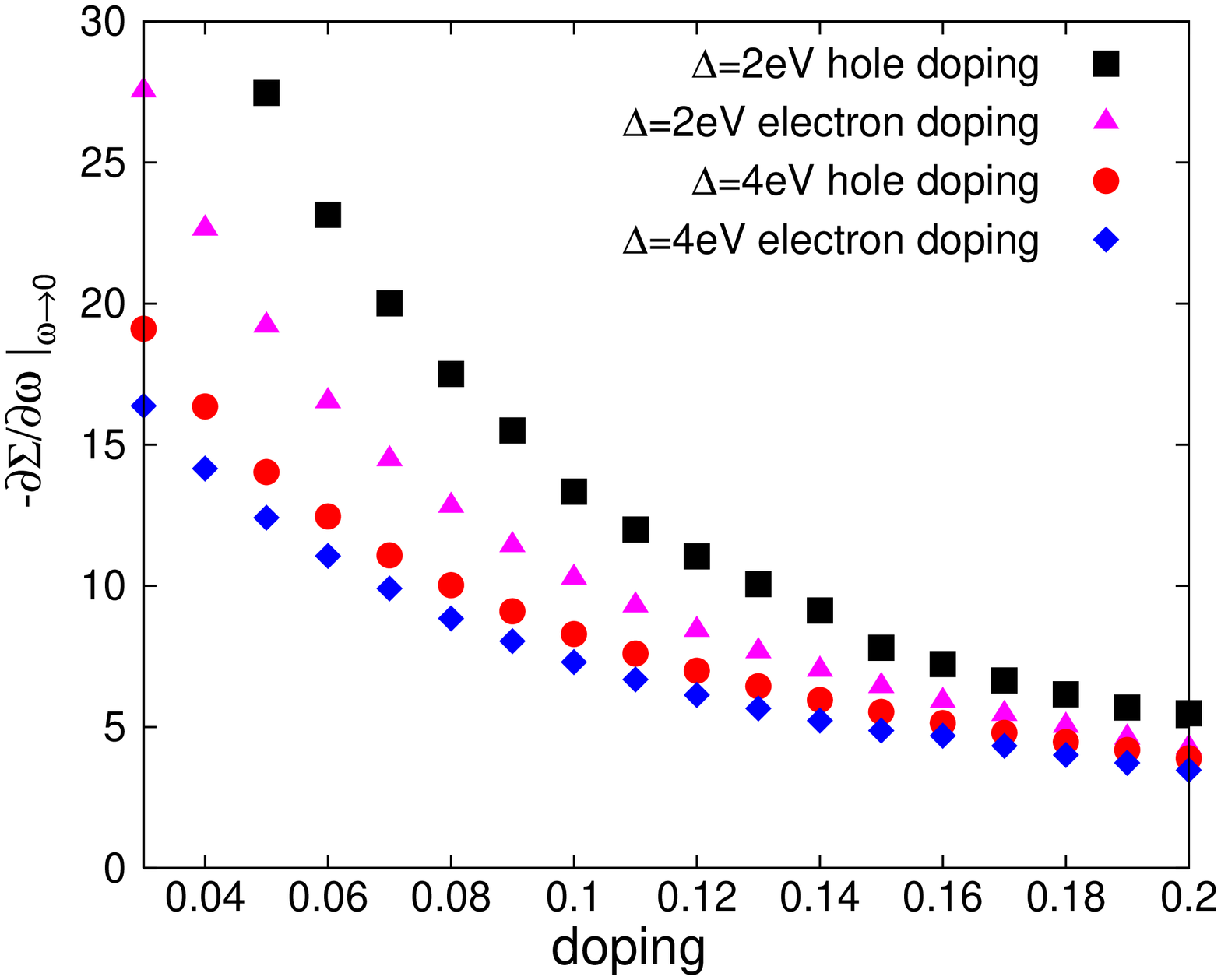}
\caption{Upper panel: comparison of CT-QMC and ED self-energies over
a wide frequency range calculated for $U=9$eV, $\Delta=4{\rm
eV}\sim\Delta_{c1}$ at dopings indicated. Lower panel: doping
dependence of  $-\partial\Sigma/\partial \omega |_{\omega\rightarrow
0}$ estimated from lowest two Matsubara points obtained from $U=9$eV
ED calculations.  } \label{selffig}
\end{figure}

To further probe the particle-hole asymmetry we show in the upper
panel of Fig.~\ref{selffig} the self-energy calculated on the
Matsubara axis for $0.15$ electron and hole doping at
$\Delta=4\mathrm{eV}\sim\Delta_{c1}$.  The near-perfect agreement
between the results of ED and CT-QMC calculations serves as a test
of the reliability of our results. The lower panel focuses on the
low frequency behavior, presenting the doping dependence of
$-\partial\Sigma/\partial\omega|_{\omega\rightarrow 0}$ estimated
from the values of ${\rm Im}\Sigma(i\omega_n)$ at the lowest two
Matsubara points of the ED calculation for both
$\Delta\lesssim\Delta_{c2}$ (paramagnetic insulator) and
$\Delta\sim\Delta_{c1}$ (paramagnetic metal).  At very low doping
the estimate becomes unreliable because the (very small) Fermi
liquid scale is not easily resolved so we do not present results. We
see again that the self-energy is systematically larger for hole
doping than for electron doping, with the difference being more
pronounced for the paramagnetic insulator case.

However, the self-energy is not necessarily the most relevant
measure of correlation strength. In a multiband model such as the
one studied here, the Fermi surface $p=p_F$ is given by the solution
of $\omega {\mathbf 1}+\mu-\mathrm{Re}{\mathbf
\Sigma}(\omega)-{\mathbf H}(p)=0$ at $\omega=0$. The degree to which
$d$ states participate in the Fermi surface states
$|\psi(p_F)\rangle$ depends on the self-energy, which of course
changes across the charge-transfer gap and with doping.  By
expanding for $\omega$ near $0$ and $p$ near $p_F$ and defining
${\mathbf V}^{\rm bare}=\partial {\mathbf  H}/\partial {\vec p}$ and
${\mathbf Z}={\mathbf 1}-\partial\mathrm{Re}{\mathbf
\Sigma}/\partial \omega$, we find that the physical quasiparticle
velocity $v^*$ is given by
\begin{equation}
v^*=\frac{\langle{\vec \psi}|{\mathbf V}^{\rm bare}|{\vec
\psi}\rangle}{\langle{\vec \psi}|\ {\mathbf Z}|{\vec \psi}\rangle}
\label{vstar}
\end{equation}
and the bare Fermi velocity by the same equation but with ${\mathbf
Z}=\mathbf{1}$. The bare velocities (defined here in physical units
by multiplying the result above by the lattice constant $3.8$\AA)
have only about $5\%$  doping dependence on either the electron on
the hole doped sides, but change substantially as one goes from
electron to hole doping: at  $\Delta=4$eV we have $v^{\rm
bare}=4.3\mathrm{eV-\AA}$ for the $0.15$ hole doped and
$3.5\mathrm{eV-\AA}$ for the $0.15$ electron doped calculation while
at $\Delta=2$eV we have $v^{\rm bare}=4.2\mathrm{eV-\AA}$ for $0.15$
hole doping and $3\mathrm{eV-\AA}$ for $0.15$ electron doping.

\begin{figure}[t]
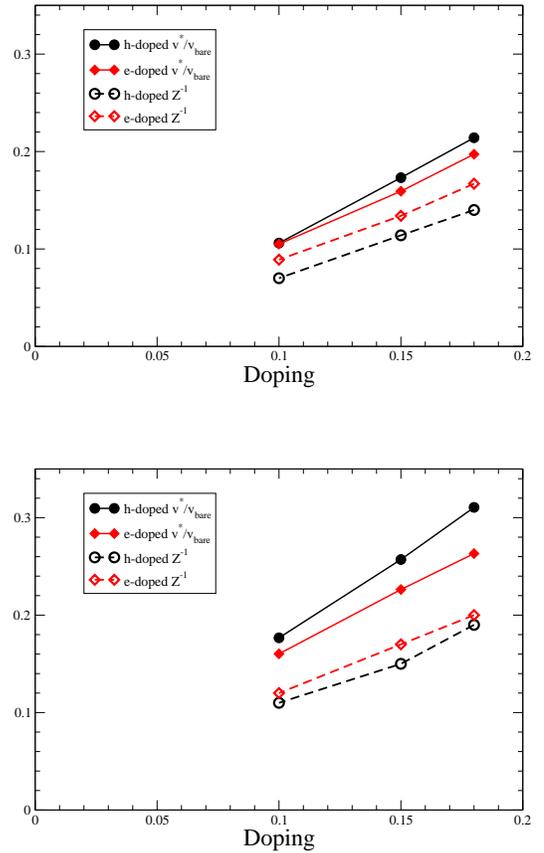

\includegraphics[angle=0,width=0.8\columnwidth]{fig4a.eps}

\vspace{0.4in}
\includegraphics[angle=0,width=0.8\columnwidth]{fig4b.eps}
\caption{Doping dependence of velocity renormalization $v^*/v^{\rm
bare}$ and inverse of self-energy derivative $Z^{-1}=1/(1-\partial
\Sigma/\partial \omega)$. Upper panel: $\Delta=2$eV; lower panel
$\Delta=4$eV.    } \label{velocityfig}
\end{figure}

We see from Fig.~\ref{velocityfig} that for the strong coupling
($\Delta=2$eV) case the particle-hole asymmetry in $\Sigma$ is
slightly overcompensated by a difference in $d$ character  of the
ground state wave function, so that the particle-hole asymmetry in
the ratio $v^*/v^{\rm bare}$ is rather  smaller in magnitude and of
opposite sign, compared to that in $\mathrm{Im}\Sigma(i\omega_n)$,
while in the more weakly correlated case the wave function changes
induce an asymmetry in velocity renormalization which, while small,
is larger than that in $\mathrm{Im}\Sigma$.

\subsection{Optical conductivity}

We have calculated the optical conductivity  and have verified the
results via comparison of the integral of our calculated
conductivity to the independently calculated ``kinetic energy'' and
to the average of the renormalized Fermi velocity over the Fermi
surface.  We also compared results obtained directly on the real
axis from the ED calculation to results obtained by analytic
continuation of the Matsubara axis self-energy obtained from CT-QMC
calculations using the methods of Ref.~\onlinecite{Wang09a}.

\begin{figure}[t]
\includegraphics[angle=-90,width=0.9\columnwidth]{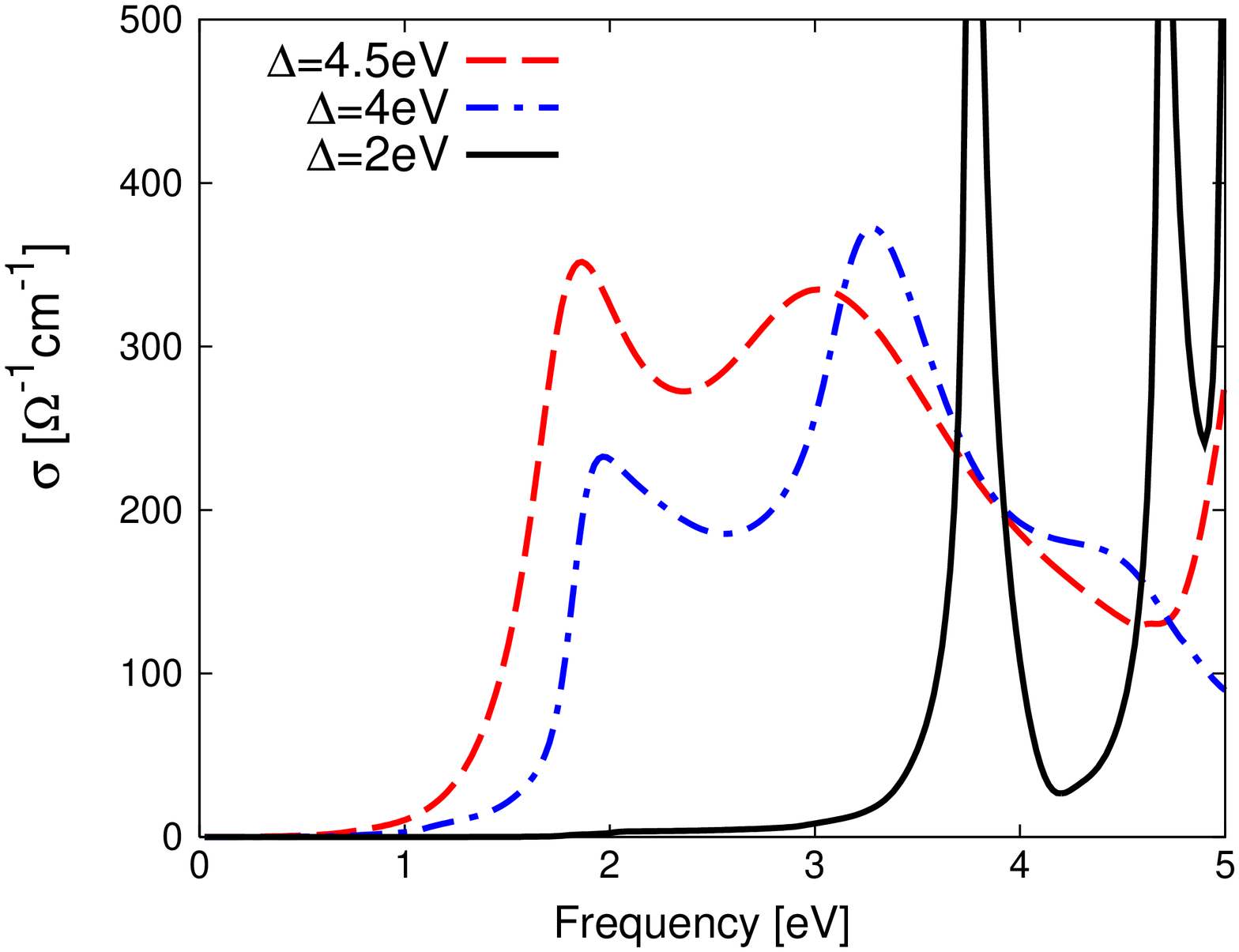}
\vspace{0.18in}
\includegraphics[angle=-90,width=0.9\columnwidth]{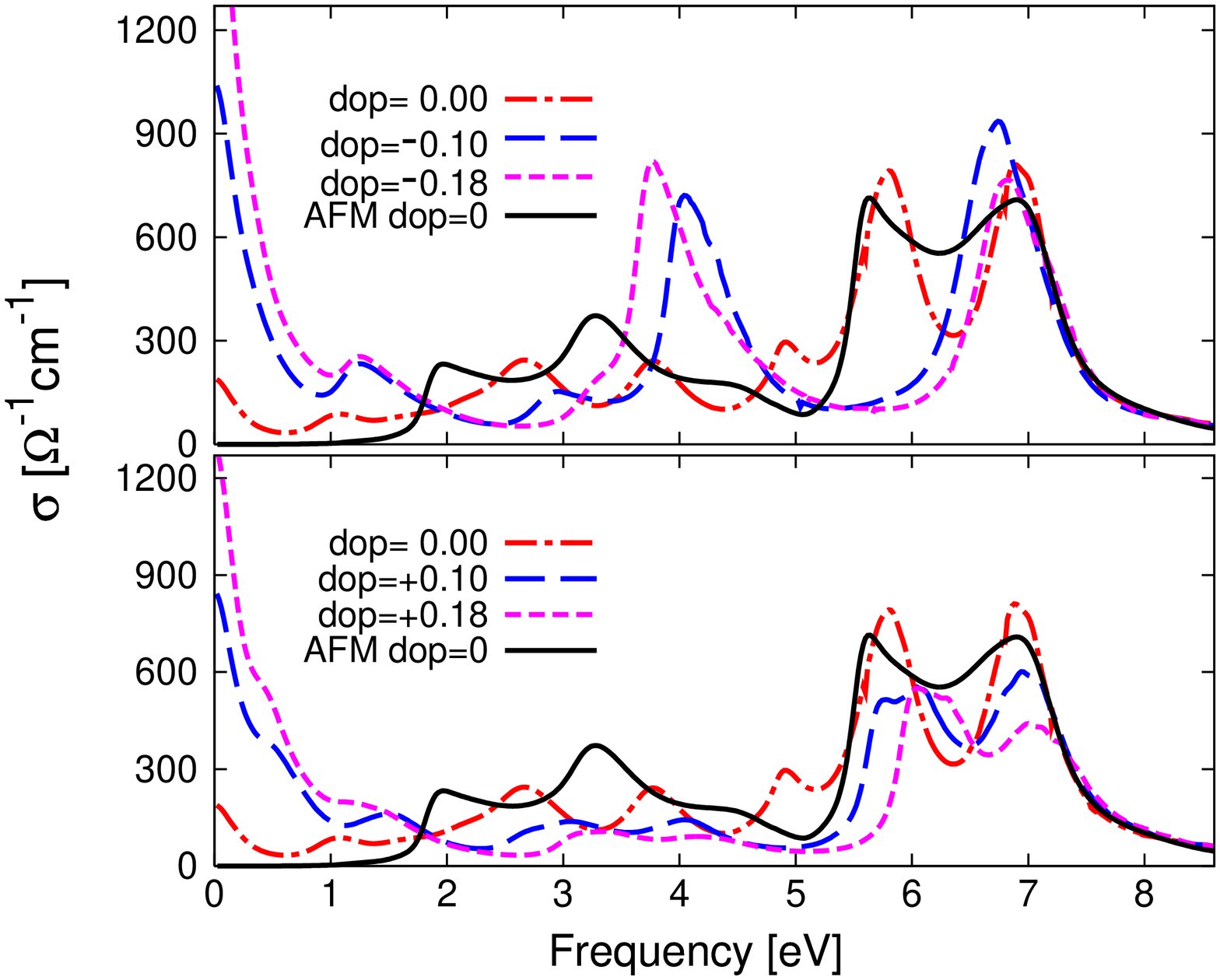}
\caption{Upper panel: optical conductivity in near-gap region
calculated for antiferromagnetic phase of  three-band model from QMC
calculation at $T=0.1$eV, $U=9$eV, carrier density of one hole per
CuO$_2$ unit (undoped case) and $\Delta$ values indicated.
Parameters: $\Delta=4.5$eV, $\varepsilon_d=-9.1$eV; $\Delta=4$eV, $\varepsilon_d=-8.8$eV; $\Delta=2$eV, $\varepsilon_d=-7.8$eV.
Lower panel: doping dependence of paramagnetic conductivity for
$\Delta=4$eV, $U=9$eV from ED calculation(red dash-dotted lines: undoped;
blue long dashed lines: $\pm0.10$ doped; magenta short dashed lines:
$\pm0.18$ doped) along with antiferromagnetic conductivity in
undoped case from QMC calculation (black solid lines, parameters are the same as in the upper panel). ED parameters:
0.18 hole doping, $\varepsilon_d=-7.7$eV;
0.10 hole doping, $\varepsilon_d=-7.9$eV;
undoped, $\varepsilon_d=-8.8$eV;
0.10 electron doping, $\varepsilon_d=-9.6$eV;
0.18 electron doping, $\varepsilon_d=-9.9$eV. } \label{condfig}
\end{figure}

The upper panel of Fig.~\ref{condfig} shows the conductivity in the
near gap region for a doping of one hole per Cu-O$_2$ unit in the
antiferromagnetic phase at $\Delta=2$eV, 4eV and 4.5eV.  As noted
above, for parameters $\Delta\sim \Delta_{c2}$ the gap in the
antiferromagnetic phase is rather larger than the experimentally
measured value $\sim1.75$eV.\cite{Uchida91} On the other hand, the
gap value determined from the antiferromagnetic phase of the $\Delta
\sim \Delta_{c1}$ calculation is in reasonable agreement with data.
This comparison places the materials clearly on the metallic side of
the paramagnetic-metal/charge-transfer-insulator phase diagram, in
agreement with previous analysis based on the one-band Hubbard
model.\cite{Comanac08} However, it is important to note that the
calculated conductivities in this frequency range are about a factor
of 2 smaller in magnitude relative to experimental data (note that a
normalization error means that  the conductivity results of
Ref.~\onlinecite{Comanac08} are too large by a factor of two). Some
of the difference may arise from transitions to bands not included
in the present calculation, but it is possible  also that the
Peierls phase arguments omit important interband matrix elements
even within the space of states we consider. This is an important
issue for further study.

The lower panels of Fig.~\ref{condfig} display the doping dependence
of the conductivity, calculated in the paramagnetic phase for
$\Delta=4$eV. Electron or hole doping adds optical absorption
strength (associated with motion of doped holes) at frequencies
$\Omega \lesssim 2$eV. For both values of $\Delta$ the integrated
absorption strength (up to 2eV) is found to be comparable for
electron and hole  doping, despite the differences in self-energy
displayed in Fig.~\ref{selffig}. In this frequency range, the
optical matrix element is proportional to the Fermi velocity which
depends on the renormalized Cu-O energy difference set by the
parameter $\Delta-{\rm Re}\Sigma(\omega)$. The shifts in ${\rm
Re}\Sigma$  as the chemical potential is tuned from hole to electron
doping leads to  a bare Fermi velocity  which is larger on the
hole-doped side than on the electron-doped side as noted in the
previous sub-section to a change in the d-character of the wave
function; these effects lead to changes in the optical matrix
elements which compensate to a considerable degree for the change in
self-energy. One of us previously argued \cite{Millis05} that the
experimentally observed similarity in low frequency optical
absorption between electron and hole-doped materials implied that
the correlation strength was about the same for two systems; we see
that this argument must be treated with caution.

A more striking change is that hole-doping but not electron doping
activates a strong transition at about 4eV between the non-bonding
oxygen band and the near-Fermi-surface states. This feature is not
observed experimentally; indeed a comparison of Figures 7 and 8 of
Ref.~\onlinecite{Uchida91} shows that  in the range  $\Omega\sim4$eV
the optical absorption in electron  doped  compounds is slightly
larger than in hole-doped compounds.  We note that in a more
realistic three-band model with oxygen-oxygen hopping included the
spectral weight in this peak would be spread over a wider energy
range. Further experimental and theoretical investigation of this
issue is important.

\section{Comparison to One Band Model}\label{comparison}

\begin{figure}[t]
\includegraphics[angle=-90,width=0.9\columnwidth]{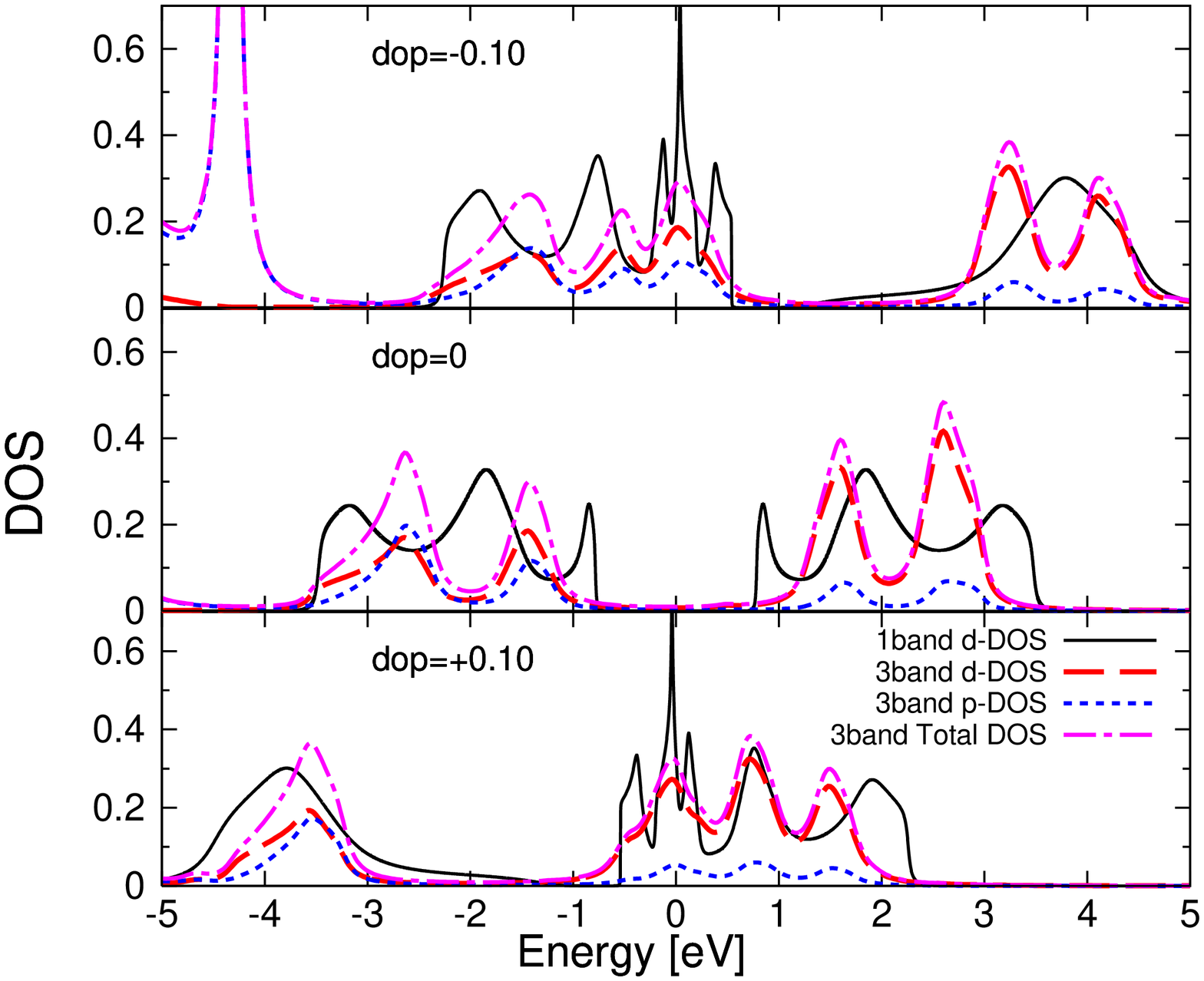}
\vspace{0.18in}
\includegraphics[angle=-90,width=0.9\columnwidth]{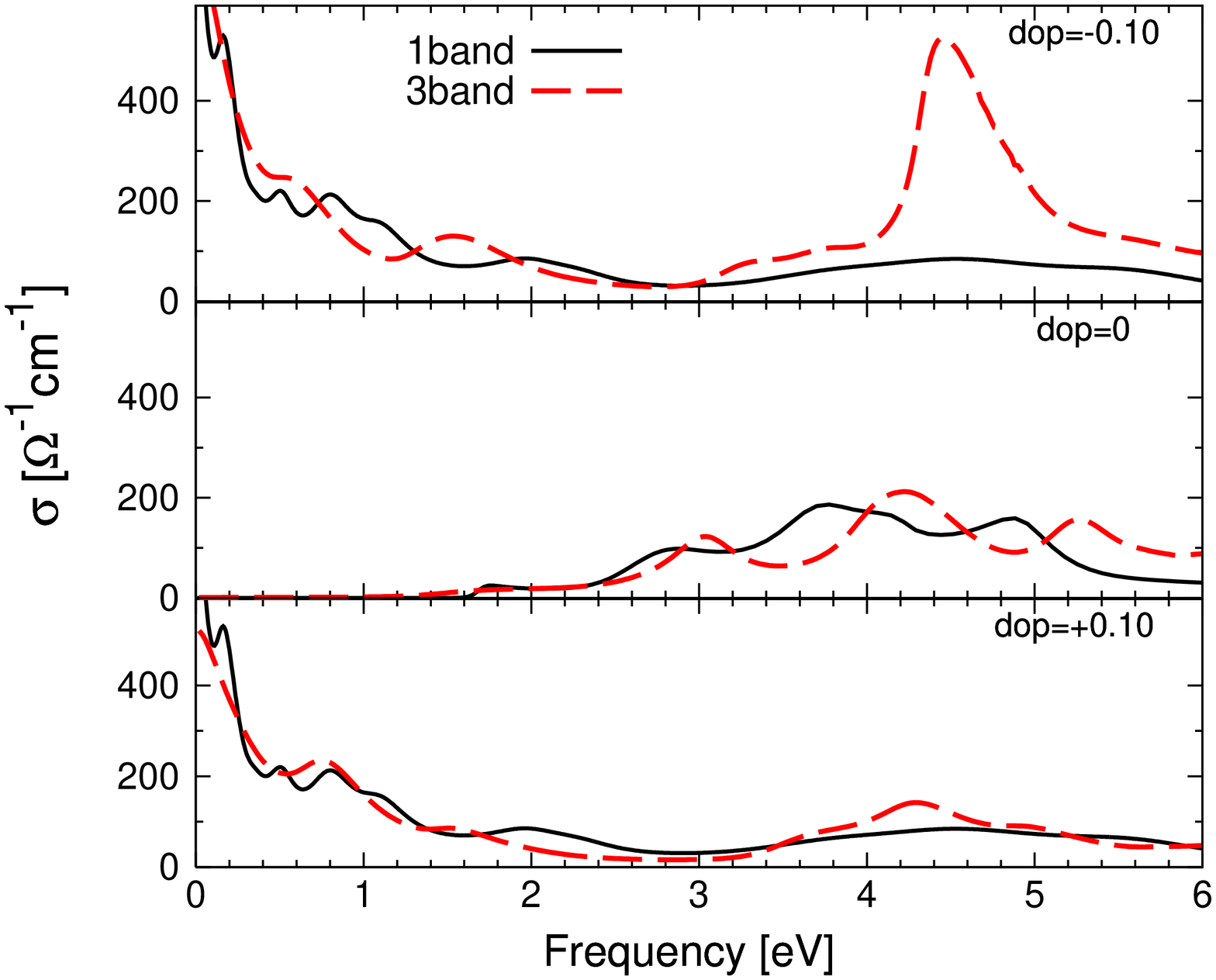}
\caption{Comparison between the three-band model and the one-band
Hubbard model at dopings indicated. Only frequencies and energies
relevant to one-band model are shown. The behavior of three-band
model at other frequencies and energies are similar to
Fig.~\ref{dos} and the lower panel of Fig.~\ref{condfig}. Upper
panel: Spectral functions. Lower panel: Optical conductivities. The
one-band model is computed at nearest neighbor hopping $t=0.37$eV,
$U_{\rm eff}=12t=4.44$eV, $T=0.1t=0.037$eV using CT-QMC solver.
Parameters: 0.10 hole doping: $\varepsilon_d=-2.7t=-1.00$eV;
undoped: $\varepsilon_d=-6t=-2.22$eV; 0.10 electron doping:
$\varepsilon_d=-9.3t=-3.44$eV. The three-band model is computed at
$U=9$eV, $\Delta=2.5$eV, $T=0$ using ED solver. Parameters: 0.10
hole doping: $\varepsilon_d=-6.8$eV; undoped:
$\varepsilon_d=-8.4$eV; 0.10 electron doping:
$\varepsilon_d=-9.5$eV.} \label{compto1band}
\end{figure}

A basic issue in the physics of the cuprates is the reduction of the
three-band model to an effective one-band
model.\cite{Emery87,Zhang88,Mila88}  We test the reduction by
comparing our results at $\Delta\approx\Delta_{c2}$ to results
obtained by applying the QMC solver to   a one-band square-lattice
Hubbard model at $U_{\rm eff}=12t\approx U_{c2}$ and $t=0.37$eV
chosen to reproduce the splitting between the centroids of the upper
and lower Hubbard bands.  Fig.~\ref{compto1band} shows the
comparison of spectral functions and optical conductivities. To
obtain spectra for the Hubbard model we used the  analytic
continuation procedure described in Ref.~\onlinecite{Wang09a}, while
the three-band results were obtained using the ED solver at
parameters $U=9$eV, $\Delta=2.5$eV and $T=0$. All calculations are
done in the paramagnetic phase.

The upper panel of Fig.~\ref{compto1band} compares the spectral
functions. One sees immediately that the one-band and three-band
models provide a reasonably consistent account of the spectra within
a few eV of the Fermi energy. Some differences are evident. From the
upper panels one sees that the one-band model has a slightly larger
bandwidth than the three-band model (the difference is most evident
for the half-filled calculation). This difference may be interpreted
as indicating that the three-band model at
$\Delta\approx\Delta_{c2}$ is equivalent to a one-band model at a
$U_{\rm eff}$ a bit greater than $U_{c2}$.

The lower panel of Fig~\ref{compto1band} compares the
conductivities. In the half filled case the two models give a quite
consistent account of the absorption  above the
Mott-Hubbard/charge-transfer gap edge. Note that because the
lowest-lying gap excitation is not optically active, the difference
in gap values noted in the previous paragraph is not easy to see in
this panel.  The one-band and three-band calculations are done with
different methods but as can be seen, the conductivities are similar
and we have verified that the spectral weight (integral of $\sigma$
up to say $1$eV) are similar for the two models.   However, for hole
doping a new feature in the three-band model appears at
$\Omega\sim4.5$eV. This is associated with transitions from the
non-bonding oxygen band, which is of course not present in the
one-band model.

\section{Conclusion}\label{conc}

In this paper we have used the single site dynamical mean field
method to solve the three-band model believed to be relevant to the
copper-oxide materials. The method includes the physics of
Zhang-Rice singlets (doped holes reside largely on oxygen sites and
have spin opposite to the copper spin), reproduces the
characteristic features of the spectrum, and reveals (especially for
parameters in the ``charge-transfer insulator'' regime) a
particle-hole asymmetry  in the self-energy which however is largely
canceled by the difference in $d$-content of the near-Fermi surface
states, leading to very similar velocity renormalizations between
electron and hole-doped compounds.  One important implication of
this finding is that (at least for the correlation strengths we have
studied) the  fact that doped holes are ``Zhang-Rice singlets''
while doped electrons are just conventional doubly occupied sites
does not imply a significant difference in the physics and
suggesting that a reduction to an effective one-band model may be
reasonable.  The issue of reduction to a one-band model has been
considered by many previous authors; however the discussion has been
largely been couched in terms related to reduction to t-J-like
models. We prefer to directly compare spectral functions and
conductivity. We found that for $\Delta\sim \Delta_{c2}$
(paramagnetic insulator) a one-band model gave a quantitatively
accurate description of the physics at scales below about $4.5$eV
(above this scale the effects of  non-bonding oxygen states which
are not included at all in the one-band model become visible).
Similar results (not shown) are found for $\Delta \lesssim
\Delta_{c1}$. It is likely that for parameters much deeper in the
insulating phase (which we have not investigated) the ``Zhang-Rice''
effects may be more important and the reduction to a one-band model
may be more problematic.

A basic question in cuprate physics is the effective correlation
strength governing the physics of the low-energy particles
responsible for superconductivity.  We addressed this question by
calculating the gap in the insulating phase. The gap is of a
``charge-transfer'' rather than ``Mott-Hubbard'' nature, as stressed
by many previous workers. Our comparison to the one-band model shows
that the charge-transfer gap can be used to extract an effective $U$
which, when used in a one-band model, reproduces the low energy
physics reasonably well.  In contrast to Ref.~\onlinecite{Weber08}
we find that antiferromagnetism has a pronounced effect on the
magnitude of the gap in the insulating state.  We find that there is
no reasonable way to obtain a gap of the physical scale (between
$1.5$ and $2$eV) if the materials are assumed to be paramagnetic
insulating side of the phase diagram; rather they must be taken to
be more moderately correlated. This conclusion is in agreement with
previous results.\cite{Comanac08}

Three features of the calculation suggest potentially interesting
directions for future research. First, the quasiparticle band
structure depends on the renormalized $d$ level energy
$\varepsilon_d^*=\varepsilon_d+\mathrm{Re}\Sigma(\omega=0)$. This
quantity has (especially for stronger correlations) a noticeable
dependence on doping and on which side of the charge-transfer gap
the materials are on. It also changes dramatically between
antiferromagnetic and paramagnetic states. Further investigation of
the physical consequences of these changes would be useful.  This
might produce additional insight into the fundamental electronic
structure dichotomy in many transition metal oxides  between the
dramatic evidence for strongly correlated behavior in high energy
spectroscopies and the more nearly band-like behavior of the lower
energy excitations.

Concerning the conductivity, as is seen most clearly in
Fig.~\ref{compto1band}, the calculated insulating state conductivity
is only a few hundred inverse-ohm-inverse-centimeters, rather
smaller than the experimentally measured values $\sim 800-1000
\mathrm{\Omega^{-1}cm^{-1}}$.  Some part of the discrepancy may
arise from other bands, not considered in our calculation, but it
may also indicate a failure of the Peierls-phase approximation to
the conductivity. A further issue in the comparison of the
high-frequency calculated conductivity to data is the appearance,
for hole doping but not electron doping, of a strong feature
relating to transitions from the non-bonding oxygen bands to the
near-Fermi-surface states. This feature implies that the high energy
conductivity in the 4-6eV range should be greater for hole doped
than for electron doped compounds. This is not seen experimentally.
Inclusion of oxygen-oxygen hopping will spread the excess spectral
weight over a wider frequency range, perhaps mitigating the
discrepancy with experiment. Further, one must bear in mind that the
electron and hole doped materials studied experimentally have
different crystal structures, introducing further uncertainties in
the comparison. However, the issue warrants further study.

\section*{Acknowledgments}
This collaboration was begun with support from the
Columbia/Polytechnique/Science Po /Sorbonne Alliance program.  AJM
and XW are  supported by  NSF-DMR-0705847 and MC  by  MIUR PRIN
2005, Prot. 200522492 and MIUR PRIN 2007, Prot. 2007FW3MJX003,  and
LdM by the  Rutgers Center for Materials Theory, NSF-DMR-0528969 and
RTRA Triangle de la physique. We acknowledge helpful discussions
with C. Weber which led to physics insights and to  correction of an
error in the normalization of the conductivity in an earlier version
of this manuscript. We also thank N. Lin for helpful discussions and
for aid in verifying the conductivity calculations.

\end{document}